\documentclass[preprint,nofootinbib,nobibnotes,groupaddress,preprintnumbers,aps,prd]{revtex4}
\usepackage{amsfonts}
\usepackage{tipa}
\usepackage{bm,epsfig,mathrsfs,amsmath,amssymb,graphicx}
\usepackage{epsfig,amsmath,graphicx,amssymb}
\usepackage{graphicx}
\usepackage{dcolumn}
\usepackage{bm}
\usepackage{subeqnarray}
\usepackage{cases}
\usepackage[toc,title,titletoc,header]{appendix}

\date{\today}

\begin{document}
\title{Efficiency and power of minimally nonlinear irreversible heat engines with broken time-reversal
symmetry}
\author{Qin Liu$^1$ }
\author{Wei Li$^1$}
\author{Min Zhang$^1$}
\author{Jizhou He$^1$}
\author{Jianhui Wang$^{1,2}$}\email{wangjianhui@ncu.edu.cn}

\affiliation{ $^1\,$ Department of Physics, Nanchang University,
Nanchang 330031, China \\ $^2\,$ State Key Laboratory of Theoretical
Physics, Institute of Theoretical Physics, Chinese Academy of
Sciences, Beijing 100190, China.}

\begin{abstract}
We study the minimally nonlinear irreversible heat engines in which
the time-reversal symmetry for the systems may be broken. The
expressions for the power and the efficiency are derived, in which
the effects of the nonlinear terms due to dissipations are included.
We show that, as within the linear responses,   the minimally
nonlinear irreversible heat engines enable attainment of Carnot
efficiency at positive power. We also find that the Curzon-Ahlborn
limit imposed on the efficiency at maximum power can be overcomed if
the time-reversal symmetry is broken.

 PACS number(s): 05.70.Ln

\end{abstract}

\maketitle
\date{\today}
\section{Introduction}
Heat engines as  energy converters provide a good platform for
studying
  the nature of thermodynamics, in addition to its relation with utilization of energy resources.
Exploring the efficient heat engines at large power is therefore an
issue of significance in thermodynamics.  The second law of
thermodynamics tells us that the efficiency of a heat engine working
between two heat reservoirs of constant temperatures $T_h$ and
$T_c(<T_h)$ is bounded by the Carnot efficiency $\eta_C=1-T_c/T_h$.
As the Carnot engine needs infinite time for completing a cycle and
produces null power, practically a heat engine needs to be speeded
up. Starting with Curzon and Ahlborn model \cite{Cur75}, the issue
of the efficiency at maximum power and its possible universal bounds
was intensively studied in the literature \cite{Wu04, Tu14, Liu15,
Hol16, Sei17, Bro05, Bra15, Esp09, Esp10, Wang15, Wang14, Rez06,
Ape12, Gon17, Fel03, Chen13, Bai18, Abah12}. Another increasing
interesting topic is the attainable maximum efficiency at
nonvanishing power for the heat engines and it has attracted much
attention recently \cite{Ben11, All13, Whit14, Pol15, Pro15, Joh17,
Lee17, Cam16}.

In the seminal paper \cite{Ben11} the bounds on efficiency for a
specific model of steady state heat engine with broken time-reversal
symmetry caused, for example, by an external magnetic field were
investigated. It was shown that, within the linear response regime,
this time-reversal antisymmetry can significantly boost the
performance and enable attainment of Carnot limit at nonzero power.
The performance of the steady state heat engine working in the
linear response regime, with broken time-reversal symmetry, raised
issues that deserve to be addressed. For instance, is there the
improvement of performance in cyclic heat engines induced by broken
time-reversal symmetry? Can heat engines beyond the linear response
regime allow the Carnot limit at positive power, with or without
broken time-reversal symmetry? How to identify the relations between
the power, efficiency, and unavoidable dissipations?  The
time-reversal symmetry was found to boost the performance of cyclic
heat engines in the linear responses \cite{Pro15, Bra15}. The
general relations between the efficiency, power and dissipations
were analyzed in the regimes of linear \cite{Jiang14, Bau16, Pro16}
and nonlinear \cite{Iyy18} responses. In another recent paper
\cite{Cam16}, the attainable maximum efficiency of heat engines is
studied that does not require broken time-reversal symmetry. It was
found that enhancement of specific heat via phase transition can
significantly boost the performance of a heat engine and enable the
realization of Carnot limit at nonzero power, even beyond linear
response regime.

In the present paper, we investigate the questions of whether the
maximum efficiency can approach the Carnot limit at positive power
and whether the Curzon-Ahlborn limit for the efficiency at maximum
power can be exceeded in the nonlinear response regime. We propose a
minimally nonlinear irreversible heat engine \cite{Izu12}, in which
the local dissipations are included \cite{Izu12, Izu13}, and  study
its efficiency and power for the case of broken time-reversal
symmetry. We show that the maximum efficiency can reach the Carnot
value at nonzero power and the Curzon-Ahlborn limit on the
efficiency at maximum power is overcomed in the time-reversal
antisymmetry.

\section{minimally nonlinear irreversible heat engine with Broken Time-Reversal Symmetry } \label{qdt}

The  heat engine model under consideration, which may be cyclic or
steady state and where broken time-reversal symmetry may be induced,
for instance, by interaction with an external magnetic field
$\textbf{B}$. The working substance is in contact with a hot
reservoir and a cold one of temperatures $T_h$ and $T_c (<T_h)$. In
order to describe the minimally nonlinear irreversible heat engines
in which  only a second-order nonlinear term is added in the linear
Onsager relations to describe the inevitable dissipations, we adopt
the extended Onsager relations \cite{Izu12, Izu13} with inclusion of
external field $\textbf{B}$,
\begin{equation}
J_{1}(\textbf{B})=L_{11}(\textbf{B})X_{1}+L_{12}(\textbf{B})X_{2},
\label{j1},
\end{equation}
\begin{equation}
J_{2}(\textbf{B})=L_{21}(\textbf{B})X_{1}+L_{22}(\textbf{B})X_{2}-\gamma_{h}J_{1}^{2}(\textbf{B}),
\label{j2}
\end{equation}
where  the nonlinear term $\gamma_{h}J_{1}^{2}$ denotes heat
dissipation into the hot reservoir and ${\gamma_{h}(\ge0)}$
indicates the dissipation strength.
 Noteworthy,  the time-reversal
symmetry will be broken due to the external field $\textbf{B}$,
thereby leading to the Onsager coefficients $L_{12}(\textbf{B})\neq
L_{21}(\textbf{B})$ for the heat engines under consideration, though
the Onsager-Casimir relation $L_{12}(\textbf{B})=L_{21}(-\textbf{B})
$ is satisfied. For sake of convenience, the following formula will
include the external field but without explicitly writing
$\textbf{B}$.

In the heat engine the heat flux $\dot{Q}_h$ is extracted from the
hot heat reservoir at the temperature $T_{h}$, and there must be a
certain heat current $\dot{Q}_c$ injected to the cold heat reservoir
of temperature $T_{c}$, with corresponding production of power
output $P=\dot{Q}_h+\dot{Q}_c$. Throughout the paper the dot means
the quantity per unit time for steady-state heat engines  or the
quantity divided by the cycle time duration for cyclic machines.
Since the entropy production of a steady-state or a cyclic heat
engine merely contributed from the two heat reservoirs, and its rate
thus reads
\begin{equation}
\dot{\sigma}=-\left(\frac{\dot{Q_h}}{T_{h}}+\frac{\dot{Q_{c}}}{T_{c}}\right)=-\frac{P}{T_{c}}
+\dot{Q_{h}}\left(\frac{1}{T_{c}}-\frac{1}{T_{h}}\right).
\label{sigma}
\end{equation}

Without loss of generality, the power output  $P$ can be expressed
as $P=F\dot{x}$, where $F$ is  an external force and $x$ is its
corresponding thermodynamically conjugate variable. As the entropy
production rate can be expressed in terms of the thermodynamic
fluxes $\textbf{J}$  and forces $\textbf{X}$:
$\dot{\sigma}=\textbf{J}\textbf{X}$, from Eq. (\ref{sigma}) we have
\begin{equation}
\dot{\sigma}=J_{1}X_{1}+J_{2}X_{2} \label{dots}
\end{equation}
through defining the thermodynamic fluxes $J_{1}\equiv\dot{x}$ and
$J_{2}\equiv\dot{Q_{h}}$, with conjugate affinities $X_1=F/T_c$ and
$X_{2}=1/T_{c}-1/T_{h}$.  The power output can thus be expressed as
\begin{equation}
P=-J_1X_1T_c. \label{power}
\end{equation}
Based on Eqs. (\ref{j1}) and (\ref{j2}), we can rewrite $J_2$ as
\begin{equation}
J_{2}=\frac{L_{21}}{L_{11}}J_{1}+L_{22}\left(1-\frac{L_{12}L_{21}}{L_{11}L_{22}}\right)X_{2}-\gamma_{h}J_{1}^{2}.\label{jj2}
\end{equation}
Let $J_3\equiv\dot{Q}_c$, we have $
J_{3}=\dot{Q_{h}}-P=J_{1}X_{1}T_{c}+J_{2},$ which takes the form of
\begin{equation}
J_{3}=\frac{L_{21}-L_{12}X_{2}T_{c}}{L_{11}}J_{1}+L_{22}\left(1-\frac{L_{12}L_{21}}{L_{11}L_{22}}\right)
X_{2}-\gamma_{c}J_{1}^{2},\label{hea}
\end{equation}
where $ \gamma_{c}=T_{h}/L_{11}-\gamma_{h} $ has been used and it
represents the strength of the power dissipation into the cold
reservoir.  With  consideration of Eqs. (\ref{j1}), (\ref{dots}) and
(\ref{jj2}), we find  that the Onsager coefficients must be
constrained by
 \begin{equation}
L_{11}\geq0,~
L_{22}\geq0,~L_{11}L_{22}-{L_{11}L_{22}\alpha\eta_{C}}-
{(L_{12}+L_{21})^2}/{4}+{L_{12}L_{21}\alpha\eta_{C}}\geq0,\label{ggg}
\end{equation}
due to the nonnegativity of the entropy production rate
($\dot{\sigma}\ge0)$. Here and hereafter we define
$\alpha\equiv1/(1+\gamma_c/\gamma_h)$ and take $\alpha$ rather than
$\gamma_c/\gamma_h$ as the dissipation ratio for simplicity. The
asymmetric dissipation limits $\gamma_c/\gamma_h\rightarrow\infty$
and $\gamma_c/\gamma_h\rightarrow0$ correspond to $\alpha=0$ and
$\alpha=1$, respectively. The symmetrical dissipation case when
$\gamma_h = \gamma_c$ leads to $\alpha = 1/2$. When the entropy
production rate tends to be zero ($\dot{\sigma}=0$),  we have $
L_{11}L_{22}-{L_{11}L_{22}\alpha\eta_{C}}-
{(L_{12}+L_{21})^2}/{4}+{L_{12}L_{21}\alpha\eta_{C}}=0. $


\begin{figure}[tb]
\includegraphics[width=3.4in]{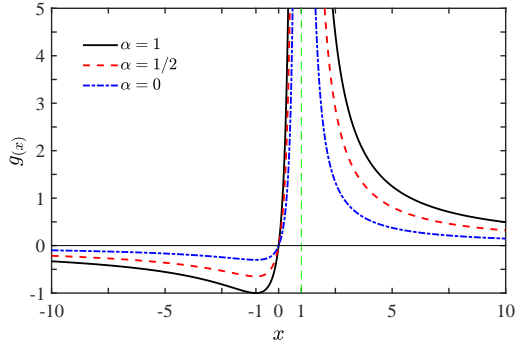}
\caption{The function $g(x)$ as a function of the asymmetry
parameter $x$, with dissipation parameter $\alpha=1$ (black solid
line), $\alpha=1/2$ (red dashed line), and $\alpha=0$ (blue
dot-dashed line). The vertical asymptote of $g(x)$ at $x=1$ is
indicated by green dotted line (When $\alpha\neq0$, $\eta_C=0.7$ is
adopted).} \label{gxx}
\end{figure}
\section{maximum efficiency }
As the efficiency $\eta$ takes the form of
\begin{equation}
\eta=\frac{P}{J_{2}}=\frac{-J_{1}X_{1}T_{c}}{L_{21}X_{1}+L_{22}X_{2}-\gamma_{h}(L_{11}X_{1}+L_{12}X_{2})^2}.\label{etam1}
 \end{equation}
the derivation of $\eta$ with respect to $X_{1}$ gives rise to the
expression of the maximum efficiency,
\begin{equation}
\eta_{\mathrm{max}}=\eta_{C}\frac{y+2-2\sqrt{y+1-\alpha\eta_{C}yx}}{4\alpha\eta_{C}+{y}/{x}},
\label{etam}
 \end{equation}
 at the thermodynamic force
\begin{equation}
    X_{1}=\frac{L_{11}X_{2}(L_{22}-L_{12}^2X_{2}\gamma_{h})-\sqrt{L_{11}(L_{11}L_{22}-L_{12}L_{21})
    X_{2}^2(L_{22}-L_{12}^2X_{2}\gamma_{h})}}{L_{11}(-L_{21}+L_{11}L_{12}X_{2}\gamma_{h})}, \label{x1h}
\end{equation}
where we have  introduced two parameters $ x={L_{12}}/{L_{21}},$ and
$ y={L_{12}L_{21}}/{(L_{11}L_{22}-L_{12}L_{21})}. $

Since no restriction is imposed on the attainable values of the
asymmetry parameter
 $x$, the relation (\ref{ggg}) yields
 \begin{subequations}
\begin{numcases}{}
    g(x) \leq y \leq 0 ~~~~~   ( x\le0), \\
    0 \leq y \leq g(x) ~~~~~( x>0),
\end{numcases} \label{gx00}
\end{subequations}
where we have defined
\begin{equation}
g(x)\equiv\frac{4(1-\alpha\eta_{C})x}{(x-1)^2}. \label{gx2}
\end{equation}
\begin{figure}[tb]
\includegraphics[width=3.4in]{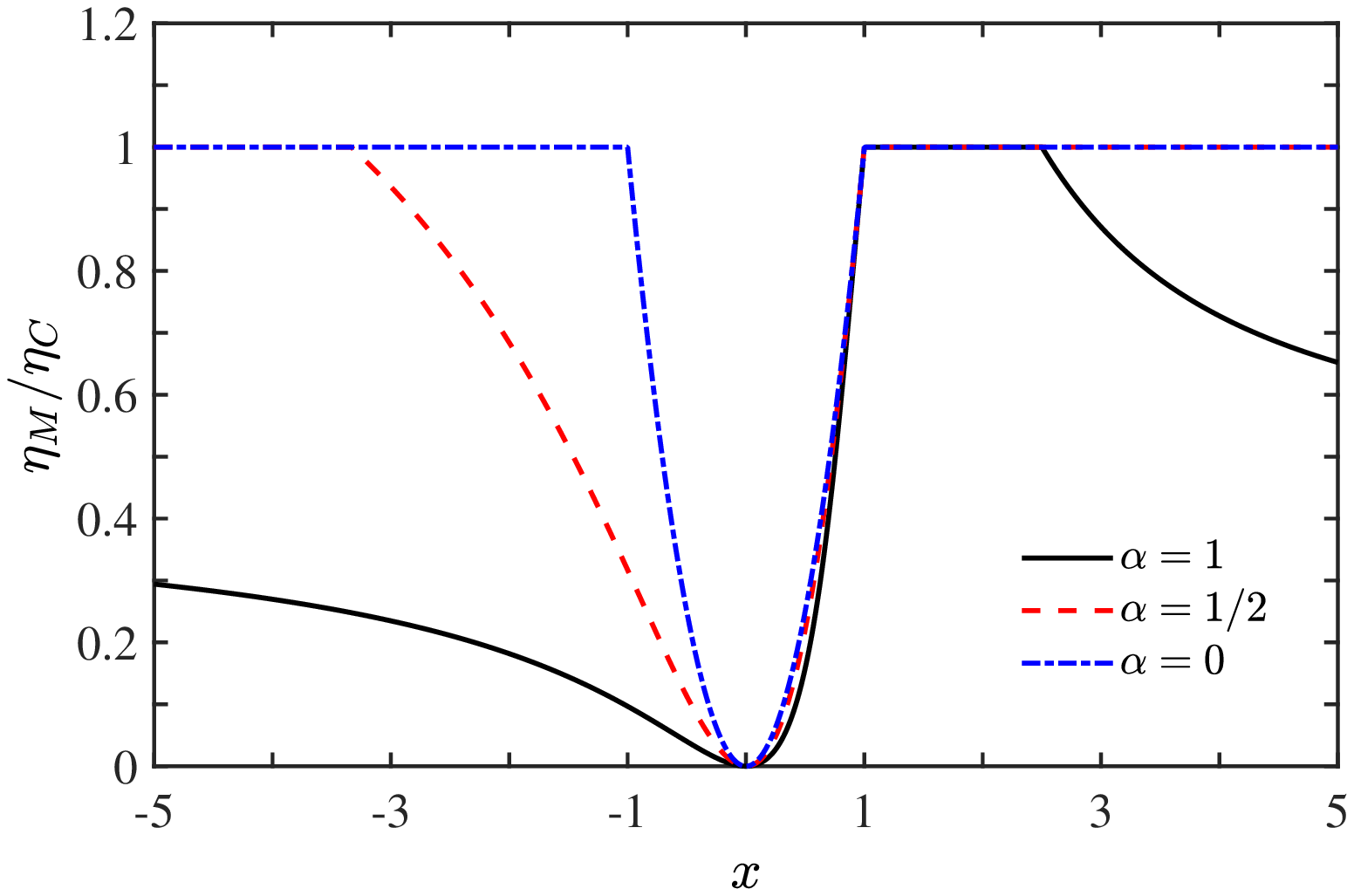}
 \caption{(Color online) Ratio $\eta_{M}/\eta_C$ as a function of the asymmetry parameter $x$.
 The dissipation ratios are $\alpha=1$ (black solid line),
  $\alpha=1/2$ (red dashed line),
 and $\alpha=0$ (blue dot-dahsed line) (when $\alpha\neq0$, $\eta_C=0.7$ is
adopted). \label{emec} }
\end{figure}
It reduces to  $g(x)=4x/(x-1)^2$ obtained in the linear response
regime \cite{Ben11}, if  the dissipation vanishes
$\gamma_h\rightarrow0$ as well as $\alpha\rightarrow 0$. We stress
that direct use of $g(x)=4x/(x-1)^2$ as done in Ref. \cite{Bai18}
would yield nonphysical,  negative entropy production rate for the
nonlinear case with $\alpha\neq0$. The effects of nonvanishing
dissipation ($\alpha\neq0)$ on the bound function $g(x)$ are of
significance for any $x$, as shown in Fig. \ref{gxx}. For a given
asymmetry parameter $x$, the maximum value $\eta_\mathrm{M}$ of Eq.
(\ref{etam}) is achieved if $y = h(x)$. Considering Eqs.
(\ref{gx00}) and (\ref{gx2}), we can obtain the maximum efficiency
$\eta_\mathrm{M}$ via simple algebra as follows: (1) when $\alpha
\eta_C\le1/2$,
  \makeatletter
 \let\@@@alph\@alph
 \def\@alph#1{\ifcase#1\or \or $'$\or $''$\fi}\makeatother
 \begin{subnumcases}
 {\eta_{\mathrm{M}}=}
 \eta_{C}\frac{x^2(1-\alpha\eta_C)}{(x-2)x\alpha\eta_C+1} &$(\frac{1}{2\alpha\eta_C-1} \leq x\leq 1)$, \label{etq1}\\
 \eta_C  &$(x\le\frac{1}{2\alpha\eta_C-1}~\mathrm{and}~ x\ge1)$, \label{ete1}
 \end{subnumcases}\label{etamx1}
 \makeatletter\let\@alph\@@@alph\makeatother
and when $1/2<\alpha\eta_C\leq 1$,
  \makeatletter
 \let\@@@alph\@alph
 \def\@alph#1{\ifcase#1\or \or $'$\or $''$\fi}\makeatother
 \begin{subnumcases}
 {\eta_{\mathrm{M}}=}
 \eta_{C}\frac{x^2(1-\alpha\eta_C)}{(x-2)x\alpha\eta_C+1} &$( x\ge\frac{1}{2\alpha\eta_C-1}~ \mathrm{and}~ x\le1)$, \label{etx2}\\
 \eta_C  &$(1\leq x\leq \frac{1}{2\alpha\eta_C-1})$. \label{etc1}
 \end{subnumcases}
 \makeatletter\let\@alph\@@@alph\makeatother
 If, in particular, $\alpha\rightarrow0$ as the dissipation vanishes $\gamma_h\rightarrow0$,
 Eqs. (\ref{ete1}) and (\ref{etq1})  simplify to
 \makeatletter
 \let\@@@alph\@alph
 \def\@alph#1{\ifcase#1\or \or $'$\or $''$\fi}\makeatother
 \begin{subnumcases}
 {\eta_{\mathrm{M}}=}
 \eta_{C}x^2 &$(\big|x\big|\le1)$, \label{etx2}\\
 \eta_C  &$(\big|x\big|\ge1)$, \label{etxe1}
 \end{subnumcases}
 \makeatletter\let\@alph\@@@alph\makeatother
which were obtained within the framework of linear irreversible
thermodynamics \cite{Ben11, Bra15}.  Besides $\eta_C$, the function
depends on $\eta_{\mathrm{M}}$ both $x$ and $\alpha$ if the
dissipation exists with $\alpha\neq0$. For $\alpha\leq
(2\eta_C)^{-1}$,  the Carnot efficiency  can be approached when
$x\ge1$ and when $x\le({2\alpha\eta_C-1})^{-1}$; whereas for
$(2\eta_C)^{-1}<\alpha\leq \eta_C^{-1}$, the range in which the
Carnot limit is reached becomes $1\leq x\leq
({2\alpha\eta_C-1})^{-1}$. The ratio $\eta_\mathrm{M}/\eta_C$ for
different values of $\alpha$ is drawn in Fig. \ref{emec}, where
$\eta_C=0.7$ for $\alpha\neq0$ is adopted. Let us  consider two
special cases: (1) when $\alpha=1$ and thus $\alpha\eta_C=0.7$, the
Carnot limit is reached during the range of $1\le x\le 2.5$, and
$\eta_\mathrm{M}=\eta_C \{3 x^2/[1+0.7(x-2)x]\}$ when $x\ge2.5$ or
$x\le1$; (2)  when $\alpha=1/2$ and $\alpha\eta_C=0.35$, the Carnot
limit is obtained in the region of $x\le-3.33$ and $x\ge1$.  The
former and latter cases are indicted by  the black solid line and
the red dashed one, respectively, in Fig. \ref{emec} where the
linear irreversible case ($\alpha=0$) is represented by the blue
dot-dashed line. Since the Carnot efficiency is obtained  under the
condition $y = g(x)$, we find that $\mathrm{det}(\mathbf{L}) =
(L_{12}-L_{21})^2 /[4(1-\alpha\eta_C)]$, and  the entropy production
rate $\dot{\sigma}=0$. The Carnot limit and $L_{12}\neq L_{21}$
yields $\mathrm{det}(\mathbf{L})>0$, showing that the Carnot
efficiency could be realized only in the non-tight coupling case.

We find from Eqs. (\ref{power}) and (\ref{x1h})  that the power  at
maximum efficiency  reads
\begin{equation}
P_{m\eta}=\eta_{C}X_{2}L_{21}^2\frac{\big|\left(x-1\right)\left[\left(1-2\alpha\eta_C\right)x+1\right]\big|
\left(\big|x-1\big|-\big|\left(1-2\alpha\eta_C\right)x+1\big|\right)^2}
{16L_{11}\left(1-\alpha\eta_C\right)^2\left(1-\alpha\eta_Cx\right)^2},\label{pmx2}
\end{equation}
which is always positive and simplifies for $0\leq \alpha\eta_C\leq
1/2$ and $1/2<\alpha\eta_C\leq 1$ to
 \makeatletter
 \let\@@@alph\@alph
 \def\@alph#1{\ifcase#1\or \or $'$\or $''$\fi}\makeatother
 \begin{subnumcases}
 {P_{m\eta}=}
 \eta_{C}x^2X_{2}L_{21}^2\frac{(1-x)[(1-2\alpha\eta_C)x+1]}{4L_{11}(1-\alpha \eta_C x)^2}
 &$(\frac{1}{2\alpha\eta_C-1} \leq x\leq 1)$, \label{pmx1}\\
\eta_{C}X_{2}L_{21}^2\frac{(x-1)[(1-2\alpha\eta_C)x+1]}{4L_{11}(1-\alpha\eta_C)^2}
&$(x\le\frac{1}{2\alpha\eta_C-1}~ \mathrm{and}~ x\ge1)$,
\label{pme1}
 \end{subnumcases}
 \makeatletter\let\@alph\@@@alph\makeatother
and
 \makeatletter
 \let\@@@alph\@alph
 \def\@alph#1{\ifcase#1\or \or $'$\or $''$\fi}\makeatother
 \begin{subnumcases}
 {P_{m\eta}=}
 \eta_{C}x^2X_{2}L_{21}^2\frac{(1-x)[(1-2\alpha\eta_C)x+1]}{4L_{11}(1-\alpha\eta_Cx)^2}
 &$(x\ge\frac{1}{2\alpha\eta_C-1} ~\mathrm{and} ~ x\le1)$, \label{pmee}\\
\eta_{C}X_{2}L_{21}^2\frac{(x-1)[(1-2\alpha\eta_C)x+1]}{4L_{11}
(1-\alpha\eta_C)^2} &$(1\leq x\leq \frac{1}{2\alpha\eta_C-1})$,
\label{pmc1}
 \end{subnumcases}
 \makeatletter\let\@alph\@@@alph\makeatother
respectively. From Eqs. (\ref{ete1}), (\ref{etc1}), (\ref{pme1}),
and (\ref{pmc1}), we see that  for $0\le\alpha\eta_C\le 1/2$ the
Carnot efficiency is attained at positive power in the range of
$x\ge1$ and $x\le(1-2\alpha\eta_C)^{-1}$, and that for
$1/2\le\alpha\eta_C\le1$ the Carnot limit can also be reached with
nonzero power if $1\leq x\leq {(2\alpha\eta_C-1)}^{-1}$. The special
case of the linear response regime when  $\alpha=0$ results into the
fact that the Carnot efficiency is achieved only when
$\big|x\big|\ge1$, as expected. We emphasize here that the nonzero
power at the Carnot efficiency is found by using $y=g(x)$, which
implies vanishing entropy production rate ($\dot{\sigma}=0$).
\section{efficiency at maximum power} \label{opi}

We now turn to the maximum power output $P_{\mathrm{max}}$ and its
corresponding efficiency $\eta_{mp}$. It follows, using Eq.
(\ref{power}) and setting $\partial P/\partial {X_1}=0$, that the
power output achieves its maximum value,
\begin{equation}
P_{\mathrm{max}}=\frac{\eta_{C}L_{12}^2}{4L_{11}}X_2,
\end{equation}
at
\begin{equation}
X_{1}=-\frac{L_{12}}{2L_{11}}{X_2}. \label{x1x2}
\end{equation}
Substituting  Eq. (\ref{x1x2}) into  Eq. (\ref{etam1}), we find that
the efficiency at maximum power is
\begin{equation}
\eta_{mp}=\frac{\eta_{C}}{2}\frac{2xy}{4+y(2-x\alpha \eta_C)},
\label{etamp}
\end{equation}
whose maximum value $\eta_{mp}^{*}$  is obtained when and only when
$y=g(x)$. By substitution of Eq. (\ref{gx00}) into Eq. (\ref{etamp})
we then arrive at
\begin{equation}
\eta_{mp}^{*}=\eta_{C}\frac{1-\alpha\eta_C}{{(\alpha\eta_C-{x^{-1}})^2}-\alpha\eta_C+1}.\label{etac1}
\end{equation}
\begin{figure}[tb]
\includegraphics[width=3.4in]{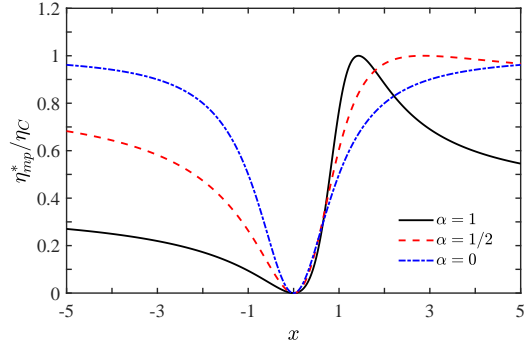}
 \caption{(Color online) Ratio $\eta_{mp}^*/\eta_C$  as a function of the asymmetry parameter $x$,
 with dissipation ratios $\alpha=1$ (black solid line), $\alpha= 1/2$
(red dashed line) and $\alpha=0$ (blue dot-dashed line) (when
$\alpha\neq0$, $\eta_C=0.7$ is adopted).} \label{etas}
\end{figure}
We note that, for $x =1/(\alpha\eta_C)$ or
$\big|x\big|\rightarrow\infty$, $\eta_{mp}^*=\eta_C$, so the Carnot
efficiency $\eta_C$ and the maximum power $P_{max}$ can be attained
simultaneously. It is therefore indicated that the limit imposed on
the efficiency at maximum power for systems with time-reversal
symmetry is overcomed  in the systems without this symmetry. If
nonlinear term vanishes ($\gamma_h\rightarrow0$ and
$\alpha\rightarrow0$), the efficiency at maximum power
$\eta_{mp}^*=\eta_{C}x^2/(1+x^2)$ in the linear situation is
recovered and $\eta_{mp}^*\rightarrow\eta_C$ as
$\big|x\big|\rightarrow\infty$. Figure \ref{etas} shows that the
efficiency at maximum power $\eta_{mp}^*$ (for given $\alpha$)
expressed by Eq. (\ref{etac1}). Insight can be gained into the
condition of attainment of the Carnot efficiency by seeing first
from Fig. \ref{etas} that for $\alpha=1$ and $\eta_C=0.7$ efficiency
at maximum power $\eta_{mp}^*=\eta_C$ can be achieved at the point
$x=1/0.7\simeq1.428$ (or $\big|x|\rightarrow\infty$ which is not
shown in the figure). Second, from Fig. 1, we note that for $x<0$
the efficiency $\eta^*_{mp}$ increase more slowly to approach the
Carnot limit in the minimally nonlinear response regime than in the
linear response case.

We also emphasize that, for the time-reversal symmetry ($x=1$),  the
efficiency at maximum power (\ref{etac1})  becomes
\begin{equation}
\eta^*_{mp}=\frac{\eta_C}{2-\alpha\eta_C},
\end{equation}
which is situated between $\eta_C/2\le \eta_{mp}^*\le
\eta_C/(2-\eta_C)$ as $0\le\alpha\le1$. The upper bounds and lower
bounds  were obtained earlier in the low-dissipation Carnot heat
engines \cite{Esp10} and the minimally nonlinear irreversible heat
engines \cite{Izu12} satisfying the tight-coupling condition  at the
asymmetrical dissipation limits. In the dissipation symmetric limit
$\alpha=1/2$, we find that the maximum efficiency at maximum power
is $\eta_{mp}^*=\eta_C/(2-\eta_C/2)$, and  its expansion in terms of
$\eta_C$ up to third order is
$\eta_{mp}^*=\eta_C/2+\eta_C^2/8+3\eta_C^3/32+\mathcal{O}(\eta_C^4)$,
which agrees well with the expansion of the famous Curzon-Ahlborn
efficiency, $\eta_{CA}=1-\sqrt{T_c/T_h}
=\eta_C/2+\eta_C^2/8+\eta_C^3/16+\mathcal{O}(\eta_C^4)$, indicating
that they have the same universality of $\eta_C/2 + \eta_C^2/8$.
\section{conclusions}
 For systems with broken
time-reversal symmetry, we have investigated the performance of
minimally nonlinear irreversible heat engines (based on these
systems).    For these nonlinear irreversible heat engines, the
maximum efficiency  can tend to be Carnot limit at nonzero power and
efficiency at maximum power can go beyond the Curzon-Ahlborn limit
when the asymmetric parameter $x$ satisfies a certain condition. Our
results hold for both cyclic heat engines and steady state ones. Our
analytical results provides a theoretical framework for
understanding of minimally nonlinear heat engines, but should also
be helpful for studying the heat devices in which higher nonlinear
terms due to dissipations are involved.

 \textbf{Acknowledgements}
We acknowledge the financial support from  NSFC(Grant Nos. 11505091,
11265010, and 11365015), the Major Program of Jiangxi Provincial NSF
(Grant No. 20161ACB21006), and the Open Project Program of State Key
Laboratory of Theoretical Physics, Institute of Theoretical Physics,
Chinese Academy of Sciences (Grant No. Y5KF241CJ1).


\end{document}